\documentclass[
    ,final            
  ]
  {aipproc}

\layoutstyle{6x9}

\begin{document}

\title{Thermal effects on the Fission Barrier of neutron-rich nuclei}



\author{Futoshi Minato}{
  address={Department of Physics, Tohoku University, Sendai 980-8578, Japan}
}

\author{Kouichi Hagino}{
  address={Department of Physics, Tohoku University, Sendai 980-8578, Japan}
}

\begin{abstract}
We discuss the fission barrier height of neutron-rich nuclei in a r-process site 
at highly excited state, which is resulted from the beta-decay or the neutron-capture processes.
We particularly investigate 
the sensitivity of the fission barrier height to the temperature, including the effect of 
pairing phase transition from superfluid to normal fluid phases. 
To this end, 
we use the finite-temperature Skyrme-Hartree-Fock-Bogolubov method with a zero-range 
pairing interaction.
We also discuss the temperature dependence of the fission decay rate.
\end{abstract}

\maketitle

\section{Introduction}
R-process is considered to be a leading candidate responsible for the synthesis of elements heavier than A$\sim$70 \cite{Wallerstein}.
Since the study of r-process involves many kinds of nuclei which extend to the neutron-rich region,
one often has to predict theoretically their masses and beta-decay rates. 
Although the recent experimental exploration to nuclei close to neutron- (proton-) drip line will make the study of r-process more reliable, a part of heavy neutron-rich nuclei produced by r-process are still difficult 
to examine experimentally.
Notice that those heavy nuclei have a characteristic decay channel, that is fission.
R-process with fission of heavy neutron-rich nuclei has been discussed 
and considered to explain the formation of elements with A$\sim$130\cite{Panov1}.

One of the important quantities which govern the fission process is a fission barrier. 
So far, the theoretical study of fission barrier height has been limited to the zero-temperature
\cite{Samyn05,Burvenich04,Stoitsov03}.
This is justified if the excitation energy is not so high and the corresponding temperature is 
much lower than the critical temperature $T_{crit}$, at which pairing correlations vanish
(the critical temperature may be approximated 
with $T_{crit}\simeq0.5\Delta_{T=0}$ \cite{Goodman1981}).
The pairing gap parameter for heavy nuclei is relatively small.
In addition, heavy neutron-rich nuclei involved in the r-process may have a high excitation energy, 
leading to the pairing phase transition, 
as a consequence of the beta-decay or the neutron-capture. 
The highly excited state may decay via fission. 
The neutron-induced fission has a significant role 
in the environment of high neutron density, while the beta-delayed fission is 
important when the neutron density is lower than about $10^{22} cm^{-3}$\cite{Panov2003}.
If the pairing phase transition occurs, the fission probability will be considerably affected. 

The aim of this work is to assess the sensitivity of fission barrier height 
and fission probability 
of hot neutron-rich nuclei 
to the temperature. We study $^{236}$U, which is close to the stability line, as well as $^{286}$Fm,  
which is neutron-rich nucleus relevant to the r-process.

\section{Framework}
In this work, we use the finite temperature Skyrme-Hartree-Fock-Bogoliubov (FTHFB) method 
\cite{Goodman1981,Sandulescu} with a zero-range pairing force.
The framework is the same as the zero-temperature HFB except for the particle density and pairing tensor.
They are modified according to the Fermi-Dirac distribution of quasi-particle occupancy. 
The thermal averaged particle density and pairing field are given as, 
\begin{equation}
\rho_T(\vec{r})=\sum_i \Big( V_i^*(\vec{r})V_i(\vec{r})(1-f_i)+U_i^*(\vec{r})U_i(\vec{r})f_i \Big),
\end{equation}
\begin{equation}
\Delta_T(\vec{r})=\frac{1}{2}V_q\sum_iU_i^*(\vec{r})V_i(\vec{r})(1-2f_i),
\end{equation}
where $f_i$ is the quasi-particle occupation probability defined by $f_i=1/(1+e^{\beta E_i})$, 
$\beta$ being $\beta=1/kT$.
We have assumed the volume-type pairing interaction,
$V_{pair}(\vec{r}-\vec{r'})= V_q \delta (\vec{r}-\vec{r'})$,
where $q=p$ corresponds to proton and $q=n$ neutron. 
$U_i$ and $V_i$ are the quasi-particle wave functions, which are the solution of 
the FTHFB equation given by, 
\begin{equation}
\left(
\begin{tabular}{cc}
$h_T-\lambda $ & $ \Delta_T $ \\
$ \Delta_T $    & $ -h_T+\lambda $
\end{tabular}
\right)
\left(
\begin{tabular}{c}
$U_i$\\
$V_i$
\end{tabular}
\right)
=E_i
\left(
\begin{tabular}{c}
$U_i$\\
$V_i$
\end{tabular}
\right),
\label{FTHFB}
\end{equation}
where $h_T$ and $\Delta_T$ are the thermal averaged mean field Hamiltonian and pairing field, respectively, 
and $E_i$ is the quasi-particle energy.

In this work, we use the quadrupole operator $\hat{Q}_2$ as a constraining operator.
For simplicity, we assume the reflection and axially symmetric nuclear shapes.
Deformation parameter used here is defined by $\beta=\sqrt{5/15\pi}\,(4\pi/3AR_0^2)\,\langle \hat{Q}_2 \rangle$,
where $R_0$ is the nuclear radius parameter given by $R_0=1.1A^{1/3}$ (fm).


%
%
\section{Results}
%
%
%
%
%

\begin{center}
\begin{figure}[htdp]
\caption{Fission barrier height as a function of temparature 
for $^{236}$U and neutron-rich nuclei $^{286}$Fm.}
\begin{tabular}{cc}
\includegraphics[height=.24\textheight,clip]{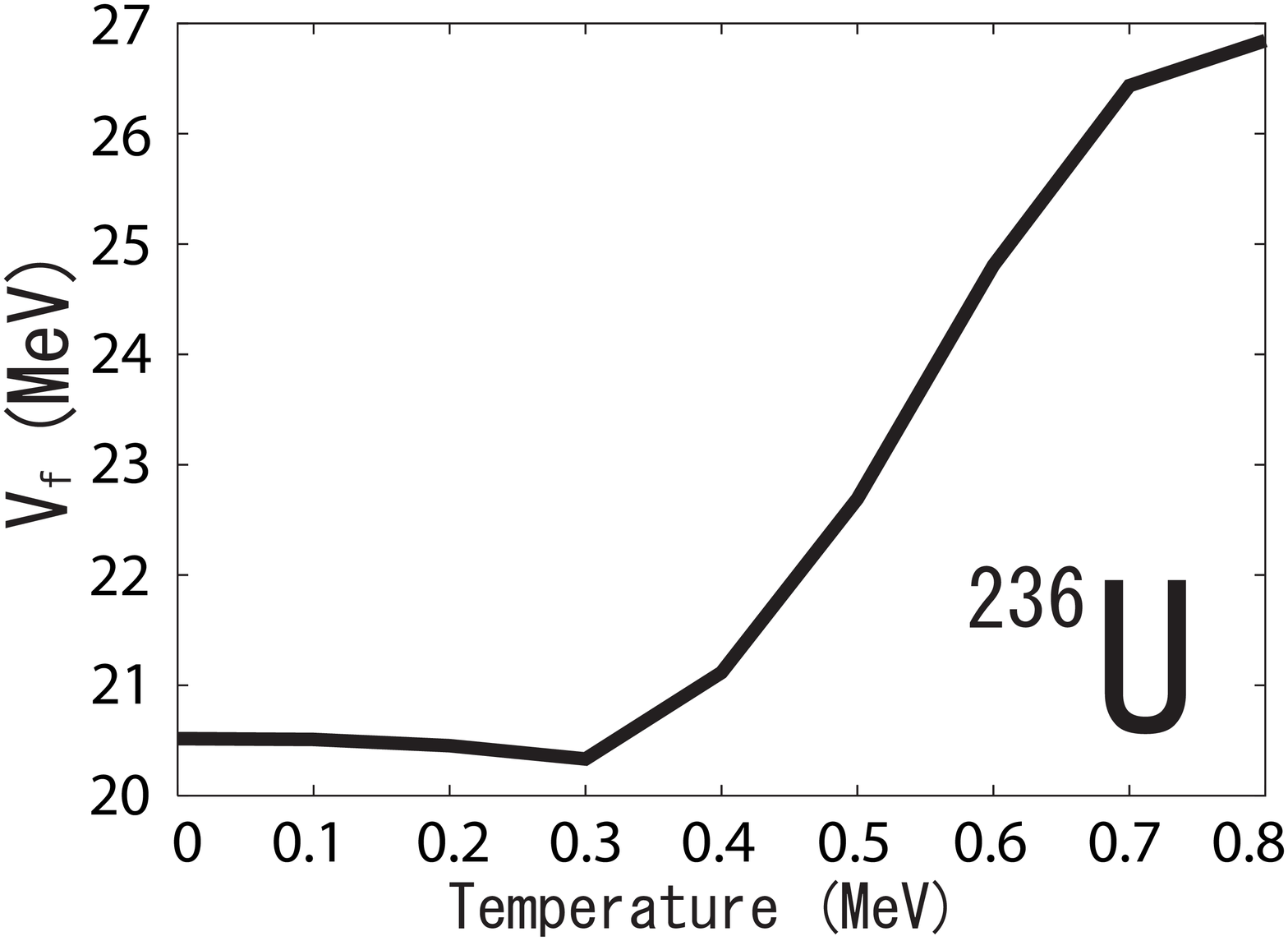}
&
\includegraphics[height=.24\textheight,clip]{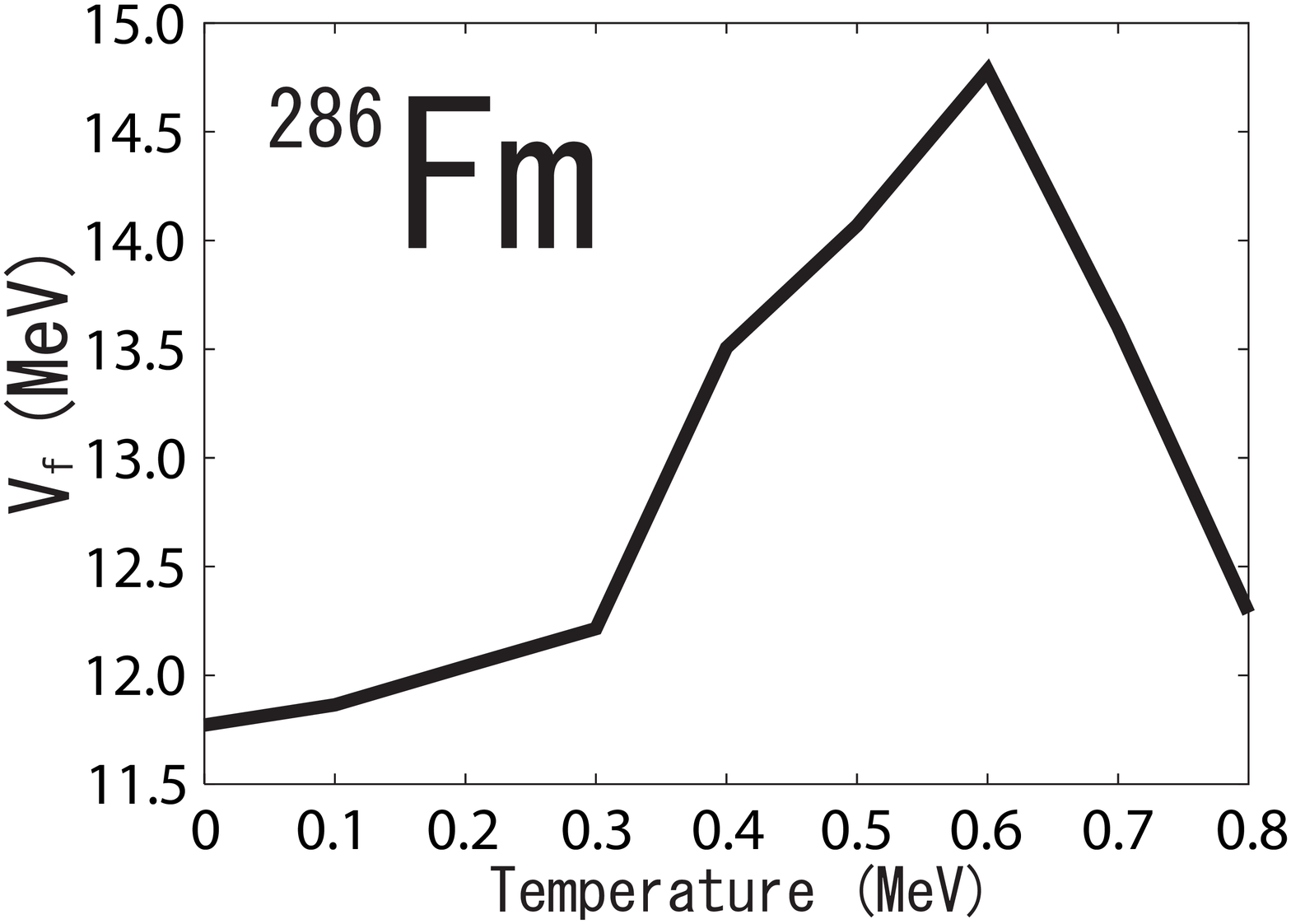}
\end{tabular}
\label{Barrier-height}
\end{figure}
\end{center}

\begin{center}
\begin{figure}[htdp]
\caption{Pairing energy at the minimum and maximum points
for nuclei $^{236}$U and $^{286}$Fm.}
\begin{tabular}{cc}
\includegraphics[height=.24\textheight,clip]{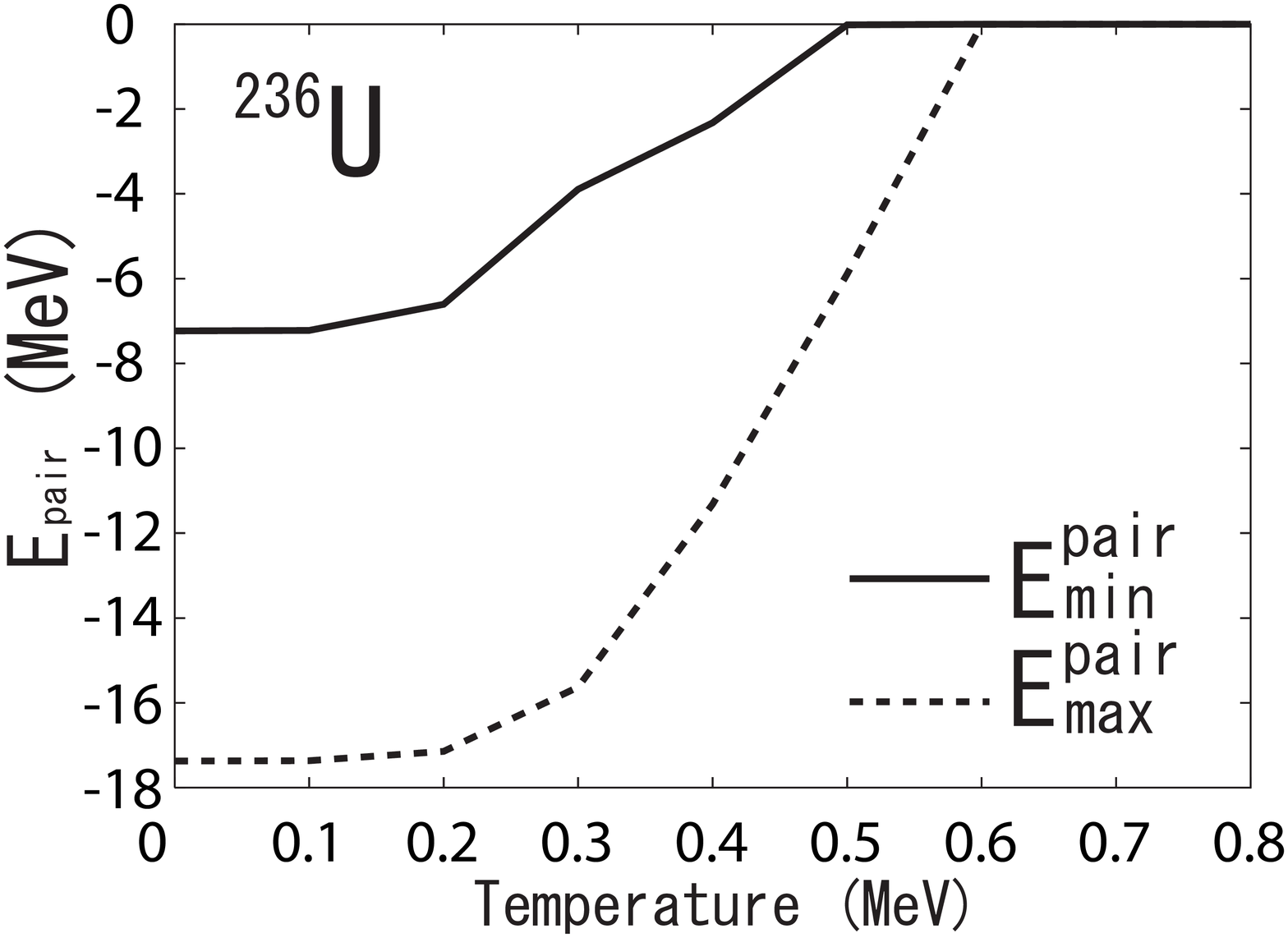}
&
\includegraphics[height=.24\textheight,clip]{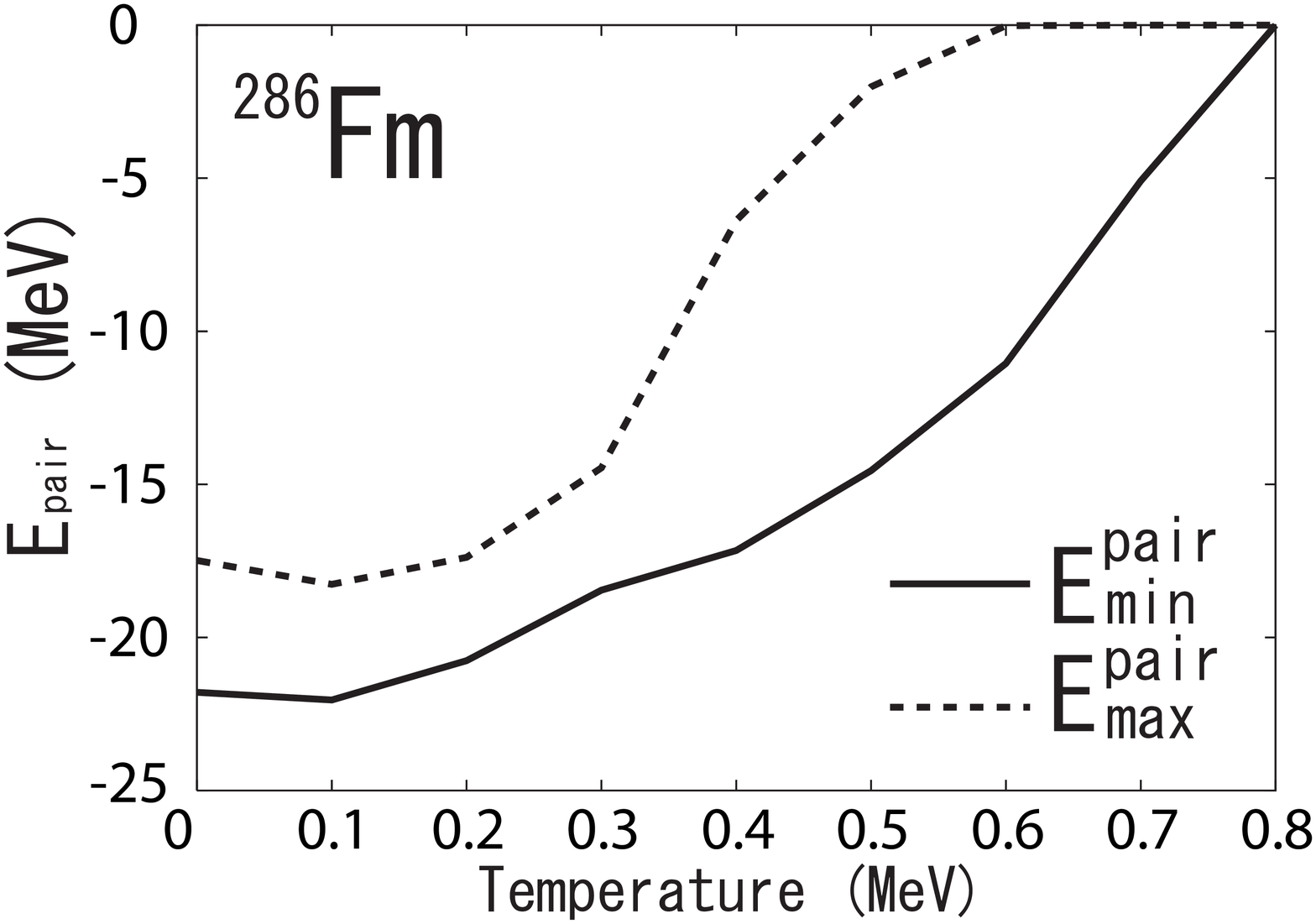}
\end{tabular}
\label{Epair}
\end{figure}
\end{center}

Figure \ref{Barrier-height} shows 
the total energy $E=\langle H \rangle$ at temperatures from $T=0.0$ to $0.8$ MeV.
We use SLy4 for the Skyrme parameter set and $V_p=-295.369$, $V_n=-286.339$ MeV for
the pairing strength, which are determined so as to reproduce the ground state energy 
of $^{236}$U.
We determine the fission barrier height from the difference between
the highest and the lowest energy in the potential energy surface.
Fig. 1 indicates that the barrier height of $^{236}$U continuously increases with temperature,
while that of $^{286}$Fm first increases but begins to decrease at T=0.6 MeV.
This feature is due to the difference in the behavior of the pairing energy 
between the two nuclei at the maximum and minimum points,
where the energy is the highest and lowest, respectively. 
The pairing energy at the minimum and maximum points as a function of temperature is
shown in Figure \ref{Epair}.
The pairing energy at the mimimum point is larger than that at the maximum point for $^{236}$U, 
while it is opposite for $^{286}$Fm. Moreover, the pairing energy at the minimum for $^{286}$Fm has 
much weaker temperature dependence than that at the maximum point. All of these features lead to 
the behaviors shown in Figure \ref{Barrier-height}.




\begin{center}
\begin{figure}[htdp]
\caption{Decay width $\Gamma_f$ for nuclei $^{236}$U and $^{286}$Fm.
The cold model uses the fission barrier at zero-temperature,
while the hot model takes into account the temperature dependence of fission barrier. } 
\begin{tabular}{cc}
\includegraphics[height=.24\textheight,clip]{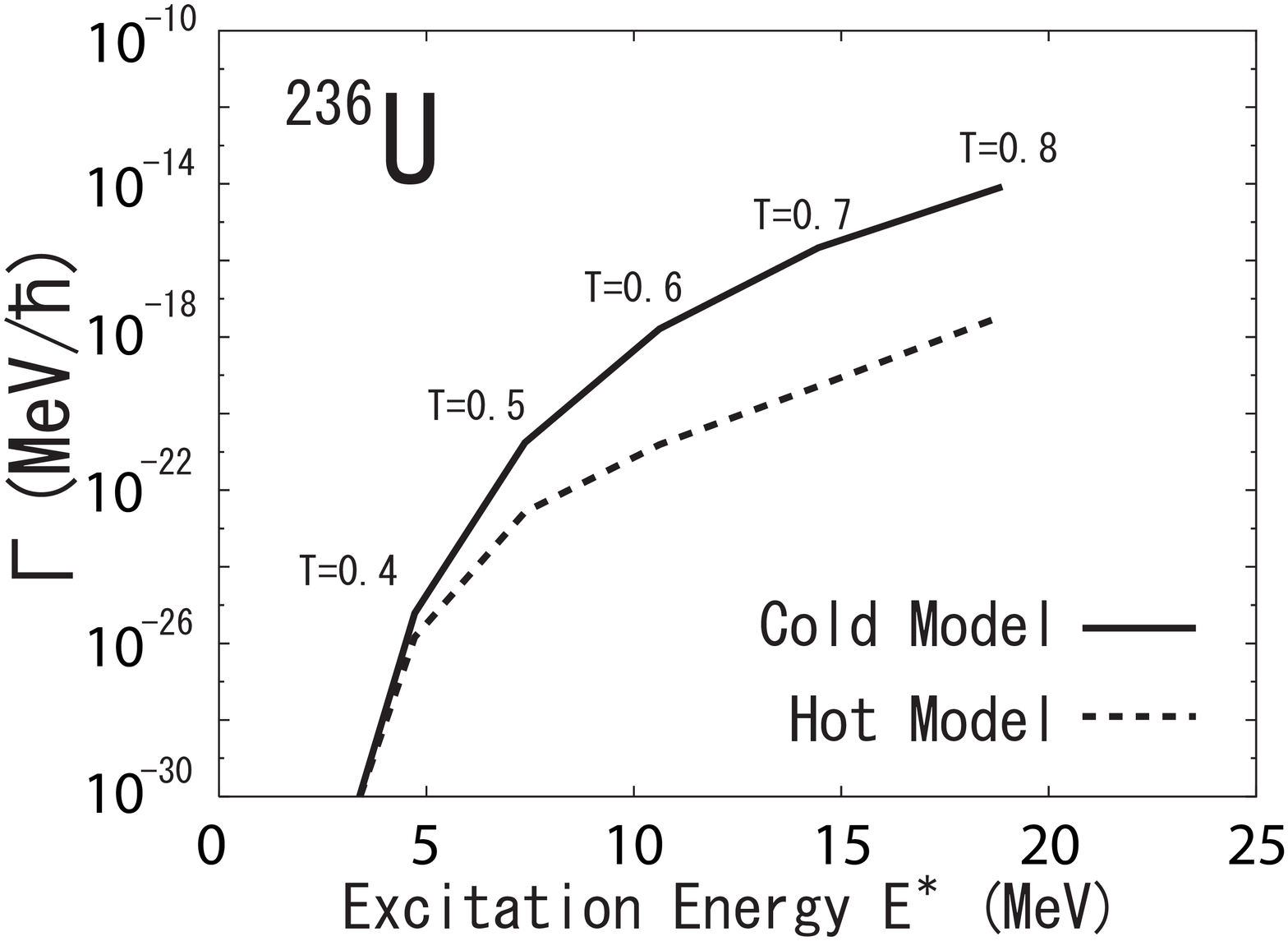}
&
\includegraphics[height=.24\textheight,clip]{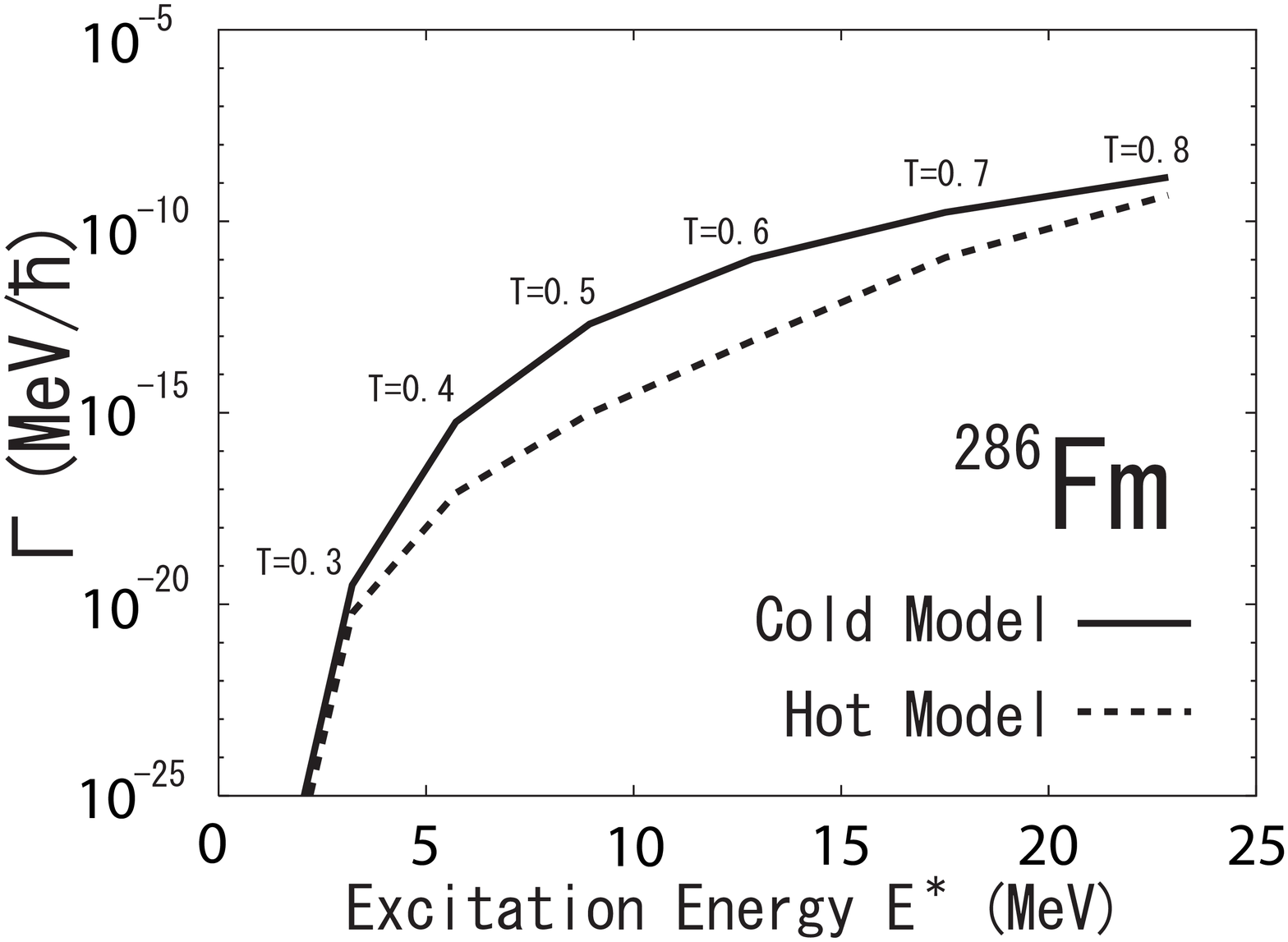}
\end{tabular}
\label{Bohr-Wheeler}
\end{figure}
\end{center}

In order to estimate the decay rate of these nuclei, 
we adopt the Bohr-Wheeler formula given by, 
\begin{equation}
\Gamma_f=\frac{\omega_0}{2\pi}e^{-\beta V_f}.
\end{equation}
The quantity $\omega_0$ is the curvature at the minimum point, 
\begin{equation}
\omega_0=\sqrt{\frac{V''(\beta_{g.s.})}{m}},
\end{equation}
where $m$ is the mass parameter.
We use the liquid-drop-model for $m$ given by $m=(3/4\pi) (1/0.04) Am_nR_0^2$
\cite{RingandSchuck}, where $m_n$ is the nucleon mass.
We show the results in Figure \ref{Bohr-Wheeler} as a function of the excitation energy, where we 
use the formula $E^*=(A/8) T^2$.
The solid line is obtained with the fission barrier at zero-temperature (the cold model) while the dashed 
line takes into account the temperature dependence of the fission barrier (the hot model). 
By using the temperature dependent fission barrier height, the decay widths decrease 
by about $10^{-1}$ for $^{286}$Fm,
and $10^{-3}$ for $^{236}$U 
at excitation energies larger than 5 MeV. 
We can see that the results of the cold and hot models for $^{286}$Fm 
become close to each other at higher temperatures due to the temperature dependence 
of fission barrier as seen in Figure \ref{Barrier-height}.

In summary, we have calculated the temperature dependence of the fission barrier for
neutron-rich nucleus $^{286}$Fm and the $^{236}$U nucleus near the
stability line.
For this purpose, we used the finite-temperature Hartree-Fock-Bogoliubov with a zero-range
pairing interaction. 
We found that the fission barrier height of $^{236}$U becomes large at a high temperature 
and that that of $^{286}$Fm shows a decrease at T=0.6 MeV
because of the temperature dependence of the pairing energy.
We used the Bohr-Wheeler formula in order to calculate the fission rate.
By taking into account the temperature dependence of fission barrier, the 
decay widths is suppressed by a facotr of $10^{-1}$ to $10^{-3}$. 

\medskip

This work was supported by the Japanese
Ministry of Education, Culture, Sports, Science and Technology
by Grant-in-Aid for Scientific Research under
the program number 19740115.

\bibliographystyle{aipproc}   

\end{document}